\documentclass{ws-ijmpa}
\usepackage{amsmath}
\usepackage{graphicx}
\usepackage{xcolor}
\usepackage{indentfirst}
\usepackage{float}
\usepackage{longtable}
\usepackage{multirow}
\usepackage{makecell}
\def\a{\alpha}

\def\p{\partial}

\def\g{\gamma}

\def\d{\delta}
\def\de{\delta}

\def\ov{\overline}
\def\ld{\lambda}

\def\e{\eta}

\def\b{\beta}

\def\a{\alpha}
\def\ast{\alpha^*}

\def\pdellx'{\frac{\partial}{\partial x'}}
\def\pdellw'{\frac{\partial}{\partial w'}}
\newcommand{\be}{\begin{equation}}
\newcommand{\ee}{\end{equation}}
\def\bed{\begin{displaymath}}
\def\eed{\end{displaymath}}
\def\bea{\begin{eqnarray}}
\def\eea{\end{eqnarray}}
\def\[{$$}
\def\]{$$}

\begin{document}

\title{A Hydrogen-Like Model for Pion and Kaon Masses \\
%Pions, Kaons, Charmonia, and Bottonium\\
Based on Dirac Quark-Hamiltonians and  
Numerical Methods} 
%\vspace{0.3in}
%\bigskip
\author{ Zhenhua Ning \\ Physics Department, University of Illinois at Urbana-Champaign,\\ 1110 West Green Street, Urbana, Illinois 61801, USA}
\author{ A. S. Davuluru, H. Raposo and Jong-Ping Hsu \\  
Department of Physics,
 University of Massachusetts Dartmouth \\
 North Dartmouth, MA 02747-2300, USA }

% Beginning of the text

\maketitle
{\small We consider a relativistic hydrogen-like model for meson mass spectra.  The model is based on a Lagrangian with new geneal Yang-Mills symmetry.  The strong interactions involves a fourth order field equation that generates a linear confining potential.  Notably, this model for meson masses requires quarks to carry only a single `confining charge'  $f_c$ rather than three color charges.   We classify meson mass spectra into sub-spectra defined by isospin, G-parity, total angular momentum and parity.  By applying a numerical solver to the Dirac Hamiltonians--incorporating two short-range confining potentials, we obtain energy eigenvalues in natural units.  The model has three coupling constants and one parameter.  It provides a robust fit of 31 pions and kaons masses (from $\approx 140 MeV$ to $\approx 2,300 MeV$) to within $\approx 18\%$  of relative percentage error.  The model suggests that the confining charge $f_c$=0.045/MeV = 8.87 fm plays the role of the basic length in the quark interactions.
  
\bigskip

%....12th draft....
%\
 %\bibitem{5, Hsu2023ws}J. P. Hsu and L. Hsu, {\em General Yang-Mills Symmetry, From Quark Confinement to an Antimatter Half-Universe}. World Scientific, 2024. xxii, chs. 8-9, Appendix C.  ???For a similar quark charge density, cf. p.132.
 
 \section{Introduction}
 % to a unified confining quark model of hadrons} 
%$  n=0, x=[4[(x+4.67)(13.55)(0.087)]^0.5 [10^4](n+0+0.75 -0.715) +4.67^2]^0.5, y=[4(6.73)(0.82)(1)(n+1/4 +1)+2.16^2]^0.5 
  
%******???
      In a previous work,\cite{Hsu2025ijmpa} we consider a helium-like model for confining three-quarks baryons based on the space-time translation $T(4)$ gaue symmetry.  The model is based on a fourth order field equation that generates a linear confining potential.\cite{GYM2024}   We use Dirac Hamitonian with approximate confining potential, so that we can solve analytically the eigenvalues with the help of Sonine-Laguerre 
 equation.\cite{Sonine1880,Hassani1879} In this paper, the model is based on a general Yang-Mills symmetry, which reduces to the usual gauge symmetry in special cases.  The pion and kaon mass spectra are obtained by solving the exact Dirac Hamiltonian through numerical approaches.
 
   For the meson mass spectra, it suffices for the model to assume only one flavor charge for quarks rather than three different color charges. For example, the pion structure in the model is denoted by $\pi(\ov{d}|u)= \pi(\text{core}|\text{shell})$, where the $\ov{d}$ quark with a larger mass is assumed to be at the core.  One crucial feature of this pion structure is that the $u$ quarks with an effective linear confining potential form an $S$ state of the shell can produce an additional effective quark Hooke potential inside the shell.\cite{Hsu2014mpla}   We note that if the effective linear potential of the $u$ quark is replaced by the Coulomb-like potential, then no additional potential is produced inside the shell.
     
  %   Since one single quark can also form an $S$ state quantum shell, similar to the electron in a hydrogen atom.  The present model for quark-antiquark mesons assumes a similar structure $\pi^+(\ov{d}|u)=\pi^+(\text{core}|\text{shell})$ for simplicity.  To calculate the approximate energy eigenstates of $\pi^+(\ov{d}|u)$, where the $u$ quark with the linear potential can produce an effective quark Hooke potential for the antiquark $\ov{d}$ at the center of the quantum shell.  
  
Based on the general Yang-Mills symmetry, we can construct the Lagrangian $L_{gYM}$ for strong quark interactions.  By imposing the strong $U_s(1)$ gauge condition $\p^\mu b_\mu=0$, we derive the fourth order field equation governing the temporal component $ b_0$ of the confining gauge field and obtain its static solution in natural units ($c=\hbar=1$) \cite{GYM2024,Hsu2025ijmpa}
\be\label{lagrangian}
L_{gYM}=\frac{f_c^2}{2} (\p^\mu b_{\mu\ld}) \p_\nu b^{\nu\ld} + i\ov{q}\g^\mu (\p_\mu+ig_c b_\mu)q - m_q\ov{q} q,
\ee
%%%%1xxx1
where $b_{\mu\ld}=\p_\mu b_\ld - \p_\ld b_\mu$, and
\be\label{scalar_field}
f_c^2 \p^2 \p^2 b^0 = g_c\ov{q}\g^0 q, \   \to \   f_c^2{\nabla^2}{\nabla^2} b_0= g_c\de^3({\bf r}),   \ \ \to  \   b_0 =\frac{-g_c}{8\pi f_c^2} r.   
\ee
%%%2
We imbed the massless confining field $b_\mu$ in the general Yang-Mills symmetric Lagrangian $L_{gYM}$, which leads to fourth-order equation for the $b_\mu$ field.  We have used the relation ${\nabla^2} (1/r)=-4\pi\de^3({\bf r}),$ to obtain the linear potential in Eq.~\eqref{scalar_field}.  It can be demonstrated\cite{Hsu6}
 that the free confining field $b_\mu$ does not propagate like the wave with a finite speed, in sharp contrast to the free massless photon.  When we solve its invariant Green's function with a source $\d^4(x)$, we see that a constant $b_\mu$ field appears instantly everywhere inside and on the forward light cone.  It indicates that the quantum of a free $b_\mu$ field is a massless particle with zero energy-momentum, just like Wigner's third class of particles.\cite{Wigner1939}

In the static limit, the zeroth component of the fourth order equation of $b^0$ in Eq.~\eqref{scalar_field} leads to the potential energy C(r) between quark and antiquark,
\be\label{cr}
 \ \   C(r)\equiv (-g_c)b_0= \left(\frac{g_c^2 }{8\pi f_c^2}\right) r ,  \ \
\ee
%%%4$$$3
 which has the units of energy MeV because the units of $f_c^2$ and $r$ are respectively $MeV^{-2}$ and $MeV^{-1}$, while $g_c$ is dimensionless in natural units.  We stress that it is not necessary to assume three different color charges for the model to give hadron masses, as demonstrated below in Table~\ref{tab:pion_full},~\ref{tab:kaon_full} and in a previous paper.\cite{Hsu2025ijmpa}

The static linear potential $C(r)$ in Eq.~\eqref{cr} is obtained by analogy with electrodynamics in the static limit and by dimensional analysis.  In quantum field theory, the linear potential is due to to the exchange of the `virtual $b_\mu$ quantum' (called b-confion) between two quarks.  Furthermore, a `virtual b-confion' appears to be physically different from a `free b-confion' in field theory.\cite{Hsu2025ijmpa,GYM2024} 

 Before using our in-house code to calculate meson mass spectra, we
  have demonstrated the ``excellent agreement'' between 
  
  (a) the analytial solution (discussed in a previous work\cite{Hsu2025ijmpa})
  for the pion spectrum based on the approxmate Hamiltonian $H_{d} \approx \a_k p_k + \b m_d +  [{(1+\b)}/{2}] [ Q r^2 + V_o],$ using the Sonine-Laguerre equation,\cite{Sonine1880,Hassani1879}  and

 (b) using our in-house code to solve the radial Dirc equations with the same approximate Hamiltonian $H_{d}$.
 
  \section{A relativistic hydrogen-like model for pions and their masses}
   
The $u$-quark in the surrounding quantum shell of a meson can produce a quark Hooke potential $V_{qH}$ inside the shell.\cite{GYM2024}  In this model, a $\bar{d}$-quark ($\bar{s}$-quark) is at the core of the shell and contributes dominantly to the masses of pions (kaons), as we shall see below.  The confining model postulates the Hamiltonian $H_{\ov{d}}$ in Eq.~\eqref{hamd} for the $\ov{d}$-quark, which moves in the quark Hooke potential generated by the $u$ quarks within the shell.\cite{}  
 
Although the linear potential in Eq.~\eqref{cr} successfully describes quark confinement, it is not by itself fully consistent with experimental observations.
High-energy experimental evidence indicates that the strong interaction in nuclear and particle physics is a short-range force. Motivated by this observation, the model introduces a concept analogous to the blackbody cavity in Planck's theory of blackbody radiation to account for the short-range nature of the strong interaction. Specifically, the exchange of the virtual quanta $b$-confions associated with the gauge field $b_\mu$ between quarks confined within a very small region is modeled as occurring inside a confining cavity. This confining cavity is presumably characterized by a fundamental length scale, $f_c \approx 9~\mathrm{fm}$.
The existence of the confining cavity modifies the quark interaction, giving rise to the short-range factor
$[{1}/(\exp(r/f_c)-1)]$,
which appears in the effective potential $C_{\mathrm{eff}}(r)$ in Eq.~\eqref{Ham_simple} and a slightly different function for the quark Hooke potential $V_{qH}$ in Eq.~\eqref{hamd}, as discussed in Appendix I. The value of $f_c \approx 9~\mathrm{fm}$ is determined below by fitting the pion and kaon mass spectra.

Thus, the model assumes the following Hamiltonians for $\ov{d}$ and $u$ in a pion, 
%and it leads to the energy eigenvalue $E_d$
 % \renewcommand\theequation{{m1}}  
 % H_{d} \approx \a_k p_k + \b m_d + ??? \frac{(1+\b)}{2}  V_{qH},   \ \ \ \    V_{qH}= Q r^2 + V_o.
 \be\label{hamd}
 H_{\ov{d}} = \a_k p_k + \b m_{\ov{d}} +  V_{qH},   \ \ \ \    V_{qH}=\frac{Q_o r^2}{exp(2r/f_c) - 1} + V_{o}.
\ee
%%%%17%%%1%%%4%%%6xxx5%%4
\be\label{Ham_simple}
H_u =  \a_k p_k + \b m_u +  C_{eff}(r),   \ \ \ \    C_{eff}(r) = \frac{Q'_o r}{exp(r/f_c) - 1}, 
\ee
where $Q_o$, $Q'_o$, $f_c$ and $V_{o}$ are parameters to be determined.
%%%%5%%%%7xxx6%%5
Note that the Hamiltonian $H_u$ for the u quark in the quantum shell has an effective short-range linear potential $C_{eff}(r)$ between the u quark and the core d quark.  Similarly, the Hamiltonian $H_d$ involves an effective short range quark Hooke potential $V_{qH}$.  The physical basis of these short range potentials is the existence of a basic distance $f_c= 8.87 fm$ in Eqs.~\eqref{cr},~\eqref{hamd},~\eqref{Ham_simple}. 

For quarks moving in a spherically symmetric potential, the Dirac Hamiltonians given in Eqs.~\eqref{hamd} and~\eqref{Ham_simple} reduce to the following radial differential equations,\cite{sakurai1967aw}
\bea\label{ham_radial}
   \frac{dF}{dr} &=&\frac{\kappa}{r}F - (E-m-V)G, \\       			 
   \frac{dG}{dr} &=& - \frac{\kappa}{r}G +(E+m-V)F\,, \nonumber
   \eea
 %%%%%6%%8xxx7xxxx6
Equation~\eqref{ham_radial} applies to both the $\bar d$- and $u$-quark Hamiltonians. For the $\bar d$ quark with mass $m = 4.67~\mathrm{MeV}$,\cite{Workman2022} Eq.~\eqref{ham_radial} is obtained from the Hamiltonian $H_{\bar d}$ in Eq.~\eqref{hamd} by taking $V = V_{qH}$. For the $u$ quark with mass $m = 2.16~\mathrm{MeV}$,\cite{Workman2022} it follows from the Hamiltonian $H_u$ in Eq.~\eqref{Ham_simple} with $V = C_{\mathrm{eff}}(r)$.

 Naturally, the pion mass $\pi^+(\ov{d}|u)$ is the sum of $E_{\ov{d}}$ and $E_u, $
 \be\label{epion}
 E(\pi) = E_{\ov{d}} + E_u, 
 \ee
 %%%%7%%%8%%%9xxx8xxxx7
 where $E_{\ov{d}}$ is the eigenvalue of $H_{\ov{d}}$ in Eq.~\eqref{hamd} and $E_u$ is the eigenvalue of $H_u$ in Eq.~\eqref{Ham_simple}.  
   
     The model is determined by using the following four independent parameters by fitting the 13 observed pion masses,\cite{Zyla2020ptep,Navas2024prd}%%% 75%
%%12%%%14xxxx9

  \be\label{const_pion}
  Q_o=2.4(10^7) MeV^3, \ \ \ \   Q'_o=\frac{g_c^2}{8\pi f_c^2}=28.29 MeV^2 ,
  \ee
  $$
    f_c={0.045}/{MeV}= 8.87 fm, \ \  V_{o} = -1785.97 MeV\,,   \ \  
  $$
  %%%%%%9  xxxxx8 
where we have used $1/MeV=197 fm$ in natural units and $g_c=1.44$ is determined by $g_c^2=8\pi f^2_c Q'_o$. 
   With these four parameters fixed, we obtain the 13 pion masses. As shown in Table~\ref{tab:pion_full}, the predicted spectrum agrees with the experimental values, with a maximum relative error of 15\% in the model.
  % with relative error [$Rel.Error \equiv ((E(\pi) − E_{data})/E_{data}]$:

{\small
\begin{longtable}{c c c p{1.0cm} c r r r p{1.0cm} r}
\caption{Calculated mass spectrum \(E(\pi)=E_{\bar{d}}+E_u\) for pion-like mesons in units of MeV, compared with experimental values. The relative error ($\delta E$) is given by  $\delta E = [E(\pi) - E(\text{Data})]/E(\text{Data}).   $}
\label{tab:pion_full} \\
\hline
$\ell$ & $\kappa$ & $I^G(J^P)$ & \makecell{pion} & $n$ & $E_{\bar{d}}$ & $E_u$ & $E(\pi)$ & \makecell{\textbf {Data}} & \makecell{$\delta$E($\mathbf{\%}$)} \\
\hline
\endfirsthead
\multicolumn{10}{c}{{\tablename\ \thetable{} -- continued from previous page}} \\
\hline
$\ell$ & $\kappa$ & $I^G(J^P)$ & Particle & $n$ & $E_{\bar{d}}$ & $E_u$ & $E(\pi)$ &\textbf{ Data} & \makecell{$\delta$E($\mathbf{\%}$)} \\
\hline
\endhead
\hline
\multicolumn{10}{r}{{Continued on next page}} \\
\endfoot
\hline
\endlastfoot

% -------- Block 1 --------
\multirow{3}{*}{0} &
\multirow{3}{*}{-1} &
\multirow{3}{*}{$1^-(0^-)$} &
\multirow{3}{*}{$\pi^+(\bar{d}|u)$} &
0 & 140.3 & 20.8 & 161.1 & $\pi(140)$ & 15 \\
 & & & & 1 & 1189.6 & 12.7 & 1202.3 & $\pi(1300)$ & -7.9 \\
 & & & & 2 & 1585.5 & 18.9 & 1604.4 & $\pi(1800)$ & -11 \\
\hline

% -------- Block 2 --------
\multirow{3}{*}{1} &
\multirow{3}{*}{1} &
\multirow{3}{*}{$1^-(0^+)$} &
\multirow{3}{*}{$a_0(\bar{d}|u)$} &
0 & 994.0 & 33.9 & 1027.9 & $a_0(980)$ & 4.9 \\
 & & & & 1 & 1481.2 & 3.2 & 1484.4 & $a_0(1450)$ & 2.4 \\
 & & & & 2 & 1924.3 & 28.8 & 1953.0 & $a_0(1950)$ & 0.16 \\
\hline

% -------- Block 3 --------
\multirow{3}{*}{1} &
\multirow{3}{*}{-2} &
\multirow{3}{*}{$1^-(1^+)$} &
\multirow{3}{*}{$a_1(\bar{d}|u)$} &
0 & 1142.3 & 15.8 & 1158.0 & $a_1(1260)$ & -8.1 \\
 & & & & 1 & 1273.3 & 5.17 & 1278.5 & $a_1(1420)$ & -10 \\
 & & & & 2 & 1627.4 & 9.0 & 1636.4 & $a_1(1640)$ & -0.22 \\
\hline

% -------- Block 4 --------
\multirow{4}{*}{2} &
\multirow{4}{*}{2} &
\multirow{4}{*}{$1^-(2^-)$} &
\multirow{4}{*}{$\pi_2(\bar{d}|u)$} &
0 & 1781.0 & 10.9 & 1792.0 & $\pi_2(1670)$ & 7.3 \\
 & & & & 1 & 1781.0 & 24.0 & 1805.0 & $\pi_2(1880)$ & -4 \\
 & & & & 2 & 1816.2 & 32.4 & 1848.6 & $\pi_2(2005)$ & -7.8 \\
 & & & & 3 & 2090.2 & 33.0 & 2123.3 & $\pi_2(2100)$ & 1.1 \\

\end{longtable}
}

   %If we consider $f_c$ as an independent free parameter, then the eigenvalues in (10-(13) can be improed.  That is, the maximum relative percent  error can be reduced from $12\%$ to $8.1\%$. ref???? [in prvious paper $Q_o=(20.28K)(10^8), Q'=20.2K^2(10^4), $ where $K=(2/202.8) MeV^3$. check again????? The reason is that the quark Hooke potyential $V_{qH}$ is determined by the basic linear potential C(r) in Eq.~\eqref{hamd} in the model.
  
\section{Mass spectra of kaons}
  Let us apply the confining model to kaons.  The spectrum of $K^+(\ov{s}|u)$ and $K^0(\ov{s}|d)$ seem relatively simple because they are all $J^P$ states with the same isospin, in contrast to the light unflavored mesons, e.g., $\pi$ mesons, etc.  However, the experimental data suggests that a better dynamical picture for the kaon system can be obtained with the following properties:

(i) The light $u$-quark in kaons forms a quantum shell, which leads to an effective quark Hooke potential $V_{qH}$ in Eq.~\eqref{hamd}, where the heavy anti-$s$-quark  ($\ov{s}$) is located at the center of the potential and provides the dominant contribution to the energy of the $K^+$.  The parity $P$ of a kaon state in the model is defined to be related to the orbital quantum number $\ell$ of the kaon state, i.e. $P=(-1)^{\ell + 1}$.\cite{Zyla2020ptep}\footnote{If the orbital angular momentum of the $q \ov{q}'$ state is $\ell$, then the parity P is $(-1)^{\ell +1}$. }

(ii) The u-quark moves in a linear potential $C_{eff}(r)$ in Eq.~\eqref{Ham_simple} and its energy contributes only a few percents to the energy (or mass) of a kaon. 

Interestingly, such a picture resembles the quantum shell of the  confining model for baryons.\cite{Hsu2025ijmpa}  In this sense, we have an approximately unified picture for the dynamics of baryons and mesons.  

The principal quantum number $n$ of a kaon state is usually defined to have the values\cite{Zyla2020ptep} 1, 2, 3..... However, in the present model, it is natural to define $n=0,1,2,...$.\footnote{ This is suggested by approximate eignvalues based on the Sonine-Laguerre equation in the previous work.\cite{Hsu2025ijmpa}  It appears that the quantum numbers $n$ is model dependent.}

% because $n$ in (C15) came from the Sonine-Laguerre equation $yd^2G/dy^2 + (A-y)dG/dy +nG=0$, where $n$ is an integer grea ter than or equal to zero.Here, the $K$ mesons are states of $q\ov{q}'$ systems, in which the parity is $(-1)^{\ell + 1}$, where $\ell$ is the angular momentum of the state.\cite{Huang1981ws, landau1951aw} 
  
Similar to Eqs.~\eqref{hamd}--\eqref{ham_radial} for pions, we have the following Dirac equations for the kaon $K(\ov{s}|u)$ with $\ov{s}$ quark in the core and the u quark in the shell
%The confining model postulates the following Hamiltonian for the $d$-quark, $H_d$, which is in the quark Hooke potential produced by the $u$ quarks in the quantum shell of the meson.  We have 
%and it leads to the energy eigenvalue $E_d$
 % \renewcommand\theequation{{m1}}  
 % H_{d} \approx \a_k p_k + \b m_d + ??? \frac{(1+\b)}{2}  V_{qH},   \ \ \ \    V_{qH}= Q r^2 + V_o.
 \be\label{hamsqk}
 H_{\ov{s}} \approx \a_k p_k + \b m_{\ov{s}} +  V_{qH},   \ \ \ \    V_{qH}=\frac{Q_o r^2}{exp(2r/f_c) - 1} + V_{o}.
\ee
%%%%17%%%1%%%12%%514xxxxxxx9
\be\label{hamuqk}
H_u \approx  \a_k p_k + \b m_u +  C_{eff}(r),   \ \ \ \    C_{eff}(r) = \frac{Q'_o r}{exp(r/f_c) - 1}, 
\ee
%%%%13%%15xxxxxxxx10
where $Q_o$, $Q'_o$, $f_c$ and $V_{o}$ are parameters to be determined.
%[[ diff equations related to the Hamiltonian with $(1+ \beta)/2$ replaced by 1:]]
%\begin{eqnarray} \label{rdirac}
In a spherically symmetric potential, the Dirac Hamiltonians given in Eqs.~\eqref{hamsqk} and~\eqref{hamuqk} reduce to the following radial differential equations,\cite{sakurai1967aw}
\be\label{ham_radial_kaon}
   \frac{dF}{dr}=\frac{\kappa}{r}F - (E-m-V)G,        			 
    \ee
 %%%%%14%%%%16xxxxxxx11
 $$ 
 \frac{dG}{dr} = - \frac{\kappa}{r}G +(E+m-V)F,\\   
 $$
%\end{eqnarray}
Equation~\eqref{ham_radial_kaon} applies to both the $\bar s$- and $u$-quark Hamiltonians. For the $\bar s$ quark with mass $m = 93.4~\mathrm{MeV}$,\cite{Workman2022} Eq.~\eqref{ham_radial_kaon} is obtained from the Hamiltonian $H_{\bar s}$ in Eq.~\eqref{hamsqk} by taking $V = V_{qH}$. For the $u$ quark with mass $m = 2.16~\mathrm{MeV}$,\cite{Workman2022} it follows from the Hamiltonian $H_u$ in Eq.~\eqref{hamuqk} with $V = C_{\mathrm{eff}}(r)$. 
  
 The mass of the kaon $K(\ov{s}|u)$ is the sum of the energy eigenvalues 
 \be\label{ekaon}
 E(K) = E_{\ov{s}} + E_u, 
 \ee
 %%%%15%%%17xxxxxxxxxxxx12(new)
 where $E_{\ov{s}} =E_{s}$ is the eigenvalue of $H_{\ov{s}}$ in Eq.~\eqref{hamsqk} and $E_u$ is the eigenvalue of $H_u$ in Eq.~\eqref{hamuqk}.  
 %In other words, since d quark and $\ov{d}$ quark have the same mass,$E_{\ov{d}} =E_d$ is the solution of (6) with $V=V_{qH}$ ,and $E_u$ is the solutin of (6) with $V=C_{eff}$.
  
 For simplicity, we use the same 4 independent constants as those for pions, as shown in Eq.~\eqref{const_pion}.  Based on Eqs.~\eqref{hamsqk}--\eqref{ekaon}, the confining model for kaons gives the following results for the masses of  $K^+(493.7)0^-, K^+(1460)0^-$, etc. in comparison with experimental values,\cite{Zyla2020ptep,Navas2024prd} as shown in Table~\ref{tab:kaon_full}.   
\begin{longtable}{c c c p{1.0cm} c r r r p{1.2cm} r}
\caption{Calculated mass spectrum \(E(\text{kaon})=E_{\bar{d}}+E_u\) for kaon mesons in units of MeV, compared with experimental values.\protect\cite{Zyla2020ptep,Navas2024prd} The relative error ($\delta E$) is given by  $\delta E = [E(\pi) - E(\text{Data})]/E(\text{Data}).$}
\label{tab:kaon_full} \\
\hline
$\ell$ & $\kappa$ & $I(J^P)$ & \makecell{kaon} & $n$ & $E_{\bar{s}}$ & $E_q$ & $E(K)$ & \makecell{\textbf{Data}} & \makecell{$\delta E$(\%)} \\
\hline
\endfirsthead
\multicolumn{10}{c}{\textbf{\tablename\ \thetable{} -- continued from previous page}} \\
\hline
$\ell$ & $\kappa$ & $I(J^P)$ & kaon & $n$ & $E_{\bar{s}}$ & $E_q$ & $E(K)$ & Data & $\delta E$ (\%) \\
\hline
\endhead
\hline
\multicolumn{10}{r}{{Continued on next page}} \\
\endfoot
\hline
\endlastfoot

% -------- Block 1: ell=0, kappa=-1, K^+ (sbar u), I=1/2(0^-) --------
\multirow{3}{*}{0} &
\multirow{3}{*}{-1} &
\multirow{3}{*}{$\frac12(0^-)$} &
\multirow{3}{*}{$K^+(\bar{s}|u)$} &
0 & 406.8 & 25.52 & 432.3 & $K^+(493.7)$ & -12.4 \\
 & & & & 1 & 1189 & 5.84 & 1194.8 & $K^+(1460)$ & -18.2 \\
 & & & & 2 & 1582.5 & 11.74 & 1597.2 & $K^+(1830)$ & -12.7 \\
\hline

% -------- Block 2: ell=0, kappa=-1, K^0 (sbar d), I=1/2(0^-) --------
\multirow{3}{*}{0} &
\multirow{3}{*}{-1} &
\multirow{3}{*}{$\frac12(0^-)$} &
\multirow{3}{*}{$K^0(\bar{s}|d)$} &
0 & 406.8 & 27.67 & 434.5 & $K^0(497.6)$ & -12.7 \\
 & & & & 1 & 1189 & 5.88 & 1194.9 & $K^0(1460)$ & -18.2 \\
 & & & & 2 & 1585.5 & 12.4 & 1598 & $K^0(1830)$ & -12.7 \\
\hline

% -------- Block 3: ell=1, kappa=-2, K^*_0 (sbar u), I=1/2(0^+) --------
\multirow{3}{*}{1} &
\multirow{3}{*}{-2} &
\multirow{3}{*}{$\frac12(0^+)$} &
\multirow{3}{*}{$K^*_0(\bar{s}|u)$} &
0 & 744.1 & 15.29 & 759.4 & $K^*_0(700)$ & 8.5 \\
 & & & & 1 & 1271 & 6.4 & 1277.5 & $K^*_0(1430)$ & -10.7 \\
 & & & & 2 & 1989 & 9.4 & 1998.5 & $K^*_0(1950)$ & 2.5 \\
\hline

% -------- Block 4: ell=0, kappa=-1, K^* (sbar u), I=1/2(1^-) --------
\multirow{4}{*}{0} &
\multirow{4}{*}{-1} &
\multirow{4}{*}{$\frac12(1^-)$} &
\multirow{4}{*}{$K^*(\bar{s}|u)$} &
0 & 885.1 & 24.2 & 909.3 & $K^*(892)$ & 1.94 \\
 & & & & 1 & 1189 & 5.74 & 1194.7 & $K^*(1410)$ & -15.3 \\
 & & & & 2 & 1585.5 & 11.73 & 1597.2 & $K^*(1684)$ & -5.2 \\
 & & & & 3 & 1837.7 & 31.2 & 1869 & --- & --- \\
\hline

% -------- Block 5: ell=1, kappa=-2, K_1 (sbar u), I=1/2(1^+) --------
\multirow{4}{*}{1} &
\multirow{4}{*}{-2} &
\multirow{4}{*}{$\frac12(1^+)$} &
\multirow{4}{*}{$K_1(\bar{s}|u)$} &
0 & 1140.4 & 15.3 & 1155.7 & $K_1(1270)$ & -9 \\
 & & & & 1 & 1271 & 6.4 & 1277.5 & $K_1(1400)$ & -8.8 \\
 & & & & 2 & 1624.7 & 9.4 & 1634.0 & $K_1(1650)$ & -1 \\
 & & & & 3 & 2132.4 & 32.5 & 2164.9 & --- & --- \\
\hline

% -------- Block 6: ell=2, kappa=2, K_2 (sbar u), I=1/2(2^-) --------
\multirow{3}{*}{2} &
\multirow{3}{*}{2} &
\multirow{3}{*}{$\frac12(2^-)$} &
\multirow{3}{*}{$K_2(\bar{s}|u)$} &
0 & 1781 & 16.9 & 1797.9 & $K_2(1770)$ & 1.6 \\
 & & & & 1 & 1781 & 21.3 & 1802.3 & $K_2(1820)$ & -0.97 \\
 & & & & 2 & 2303.8 & 25 & 2328.8 & $K_2(2250,2319)$ & 0.4 \\
\hline

\end{longtable}

   Note that the last data $K_2(2250,2319)$ in Table~\ref{tab:kaon_full} are essentially considered different historical tags or observations of the same underlying kaon state. According to the Particle Data Group, both $K_2(2250)$ and $K_2(2319)$ are overlapping observations representing the same high-mass, excited kaon resonance.\cite{Zyla2020ptep,Navas2024prd}   
       
 \section{Discussions}
   Within the framework of the usual quantum field theory, it is very difficult to understand roughly 500 different masses of mesons and baryons.  The main reason is that the usual second order field equations cannot lead to  a reasonable confining potential for the Dirac Hamiltonian to produce the observable hadron spectra.  The present situation is similar to that of the atomic structure, which challenged Bohr and other physicists 100 years ago.  Consequently, novel theoretical frameworks are required to explore and characterize complex hadron spectra. 
 
Fortunately, we have a reasonable quark model proposed by Gell-Mann and Zweig.  Our model for understanding hadron mass spectra is guided by the new general Yang-Mills symmetry.\cite{GYM2024}  Based on the strong $U_s(1)$ group, the model has the fourth-order gauge field equations and the confining linear potentials in the static limit.  The model for mesons assumes the specific structure that a heavy quark at the core and a lighter quark (or even the heavy antiquark with the same  mass) forming a surrounding quantum cloud.  It is similar to the understanding of hydrogen energy spectra based on the Dirac equation with the electromanetic U(1) gauge symmetry.  Furthermore, we introduce the concept of ``confining cavity" for the interactions of quarks and the very strange b-quanta, where a free b-quantum is massless and have zero energy-momentum.\cite{GYM2024}  As a result, the linear confining potential becomes effectively short ranged, so that it is  consistent with experiments of strong interactions.  

In this connection, it is interesting to compare the pion masses calculated on the basis of the potential with and without the confining cavity.  The results for the pion spectrum with $\ell=0$ and $\kappa=-1$  are summarized in the table below, which corresponds to $\pi^+(\bar{d}|u)$ entry in Table~\ref{tab:pion_full}.
%\newpage
\begin{table}[!ht]
\centering
\caption{Comparison of the pion energies $E(\pi)$ without and with the cavity.}
\label{tab:pion_cavity}
\begin{tabular}{c|c|c|c}
\hline
$n$ & $E(\pi,\text{ without cavity})$ & $E(\pi,\text{ with cavity})$ & \textbf{Data} \\
\hline
0 & 142   & 156   & $\pi(140)$ \\
1 & 436   & 1197  & $\pi(1300)$ \\
2 & 709   & 1599.4 & $\pi(1800)$ \\
\hline
\end{tabular}
\end{table}
This suggests that the short range property of the quark force is necessary for understanding the meson and baryon spectra.
   
  In general, the model has a maximum average percent deviation $\approx 18\%$, and most meson masses have about $\pm 8 \%$ or smaller, as shown in Table~\ref{tab:pion_full} and Table~\ref{tab:kaon_full}.  It is interesting to note that these meson mass spectra do not have a simple pattern, in sharp contrast to that of the hydrogen energy spectrum.  Furthermore, in the absence of the short range function $ 1/[{\exp(r/f_c) - 1}]$ in the confining potential, the numerical solutions give a very large maximun average percent deviation, $\approx 66\%$, as shown in the pion mass $\pi(1300)$ above.
  % in Table~\ref{tab:kaon_full}.\textcolor{blue}{I don't see the 50\% deviation. Which one it refers to?}
  
  The maximum average deviation $\approx 18\%$ in the model can be reduced to about $15\%$, if the number of prameters increases.  For example, if we use two different $V_o$ for pions and kaons, i.e., $V_o=-1785.95 MeV$ for pions and $V_o= -1686.23$ for kaons, the the maximum average deviation will be smaller, i.e. $\approx 12\%$.
  
  The numerical solutions of the Dirac equation in Eqs.~\eqref{ham_radial} and~\eqref{ham_radial_kaon} may not be trivial.  We used a fortran program to obtain the meson mass spectra in Table~\ref{tab:pion_full} and Table~\ref{tab:kaon_full}, as discussed in Appendix II.  Suppose one use AI or NDEigensystem to solve the Dirac equations, The two Dirichlet conditions are specifying that the functions go to zero at the left and right sides of the boundary. One may not obtain reliable answers because one may have numerical instabilities in the solution.  To resolve these instabilities, one must make sure that the source code can count the nodes of the wave funcitons.  For example, in our model, for a given n and $\ell$, the number of nodes for a function should be n. Otherwise, one cannot obtain reliable solutions.  Moreover, Eqs.~\eqref{ham_radial} and~\eqref{ham_radial_kaon} contain a singular term $\kappa/r$ that diverges at $r=0$, unless this can be handled internally, one cannot have the solution in the range from  0 to $r_{max}.$
 
   It is usually speculated that the hadron masses are due to the relativistic motion of quarks (and gluons) in the usual QCD. Nevertheless, the confining  model suggests a new mass generation mechanism.\cite{Hsu2025ijmpa} Namely, once the core $d$ quark comes to the center of the quark Hooke potential, its mass instantaneously increases from $\approx  4$ MeV to $\approx 140MeV $ or even to $\approx 2000 MeV$ as shown in Table~\ref{tab:pion_full}.  Thus, the quark Hooke potential $V_{qH}$ generated from the quark in the quantum cloud is crucial for the model to generate large masses of hadrons.  Such a new property of particle mass generation is interesting because of the new general Yang-Mills symmetry.\cite{GYM2024}   
    
    We  stress that the new efficient mass generation mechanism is intimately related to the large coupling constant $Q_o\approx 10^7 MeV^3$ in the quark Hooke potential in Eq.~\eqref{hamd}.  Moreover, the constant of integration $V_o$ in the quark Hooke potential in Eq.~\eqref{hamd} also plays an important role in our understanding of the meson mass spectra.  It appears that the role of $V_o$ in the model is similar to that of the vacuum expectation value of the Higgs field in Weinberg's unified electroweak theory of leptons with the spontaneous symmetry breaking.   
        
    In conclusion, the 31 meson masses in Table~\ref{tab:pion_full} and Table~\ref{tab:kaon_full} with irregular mass gaps are  based on two confining potentials with four constants.  This simplicity appears to sugget that the quark model with general Yang-Mills symmetry point to the right direction for understanding hadron spectra.  Furthermore, the predicitons for the next excited $K^*(1869)$ and $K_1(2165)$ states with n=3 are listed in Table~\ref{tab:kaon_full}, which can be used to test this confining model in the future.   
 
The work was supported in part by the Jing Shin Research Fund and Prof. Leung Memorial Fund, UMass Dartmouth Foundation.
\bigskip

\noindent
\appendix

\section{Effective short-range linear potntial in a model with `quntum cavity' for quark interactions}
 \bigskip 
 
Suppose a confined quark inside a hadron could be pictured as a particle in a `quantum cavity,' similar to atoms in the blackbody radiations.\cite{sakurai1967aw}   The model assumes that a quark and b-quantum interact in a quantum cavity with equilibrium.\footnote{Leon Hsu, private correspondence.  A free massless b-quantum satisfies a fourth-order field equation and has zero energy-momentum.\cite{Hsu6} See also Hsu and Hsu, {\em Broad Relativity}, Appendices, World Scientific, forthcoming.}  A confining quark is assumed to correspond to a set of virtual quarks in a `cavity'.   Namely, virtual quarks $\a$ and $\b$ can freely exchange a `linear potential energy' by the reversible process, e.g.,
 \be
    \a \to \b + b, \ \ \ \   \b+b \to \a,
  \ee
  %%%%33....A1
where the virtual b is the quantum of the confining field $b_\mu$ described by the Lagrangian in Eq.~\eqref{lagrangian} with general Yang-Mills $U_s(1)$ symmetry.\cite{GYM2024}   The model assumes a `confining cavity', in which the populations of virtual $\a$ and $\b$ quarks are denoted by $P(\a)=\exp(-r_\a/f_c)$ and $P(\b)=\exp(-r_\b/f_c)$ in spherical coordinate, where $r_\a (r_\b)$ is the $\a (\b)$ quark's radial coordinate.  Suppose an equilibrium is established, one has the relation,  
  \be
  P(\a) W_{\text{emis}} =P(\b) W_{\text{abs}},
  \ee
  %%%%34....A2
where $W_{\text{emis}}$ and $W_{\text{abs}}$ are respectively the transition probabilities for $\a \to \b+C$ and $\b+C \to \a$. Thus, the confining model with a set of confining cavities has effectively 
\be
  \frac{P(\a)}{P(\b)} = \frac{\exp(-r_\a/f_c)}{\exp(-r_\b/f_c)}= \exp(r/f_c), \ \ \ \   
  \ee
  %%%%35...A3
  where $r=r_\b - r_\a$, where $f_c$ is postulatd to be interpreted as a basic length for quark interactions.
  
 Based on the Hermiticity of the interaction Hamiltonian in the strong U(1) gauge invariant Lagrangian in Eq.~\eqref{lagrangian}, the model has
  \be
  \frac{W_{\text{emis}}}{W_{\text{abs}}} = \frac{n_r +1}{n_r},
\ee
%%536%%%%%36....A4
where $n_r$ corresponds the number of photons in the cavity of the blackbody radiations.\cite{sakurai1967aw}   The relations (A2)-(A4) lead to
\be
n_r= \frac{1}{\exp(r/f_c) - 1}.
\ee
%%%%37....A5
Thus, the model has an effective confining potential $C_{\text{eff}}(r)$, which has a short-range  property,
\be
C_{\text{eff}}(r)= C(r) n_r= \frac{Q'_o r}{[\exp(r/f_c) -1]},   \  \ \  \ \ \ \    Q'_o=\frac{g^2_c}{8\pi f^2_c},
\ee %%%%%%%12%%%14%%%38....A6
where the basic length $f_c=8.87 fm$ is  the strong coupling strength $f_c$.  This interpretation turnes out to be supported by the agreement with the meson mass spectra, as we shall see in Table~\ref{tab:pion_full} and  Table~\ref{tab:kaon_full}.
% to be determined by the data of mass spetra, see (8)-(11), and (16)-(22), etc.  [[Is it a length for a sub-spectrum or for all mass spectra.???]]

Suppose the quark in the quantum shell has the linear potential $C_{eff}(r)$ in (A6).  It can be shown that it can produce a new quark Hooke potential $V_{qH}$ inside the quantum shell
\be
V_{qH}(r)=\frac{Q_o r^2}{exp(2r/f_c) -1} + V_{o},
\ee
%%%%%%%%13%%%%15%%%%39....A7
where $V_o$ is a constant of integration and a small approximation is made.
%\be
%Q_o = 3.3(10^7) MeV^3, \ \ \  f_c=\frac{0.045}{MeV} = 8.87 fm,  \ \  V_{o}\approx  -2704 MeV.-1686.23???
%\ee
%%%%40...A8
 %In (A8)-(A9), there are three parameters $K$, b and $f_c$, which could be adjusted to get a good fit (say within 10 percent  diviations.)???  [[Is it a length for a sub-spectrum or for all mass spectra.???]]

\section{Numerical Solution of the Coupled Radial Dirac Equations}
The coupled radial Dirac equations for a quark in a central potential are given by
\begin{align}
    \frac{dP_{n\kappa}}{dr} &= -\frac{\kappa}{r}P_{n\kappa} + \big[E + m_d - V(r)\big]Q_{n\kappa}, \label{eq:dpdr} \\
    \frac{dQ_{n\kappa}}{dr} &= \frac{\kappa}{r}Q_{n\kappa} - \big[E - m_d - V(r)\big]P_{n\kappa}, \label{eq:dqdr}
\end{align}
where $P_{n\kappa}(r)$ and $Q_{n\kappa}(r)$ are the upper and lower radial components, $E$ is the energy eigenvalue, $m_d$ is the quark mass, $\kappa$ is the relativistic quantum number,\footnote{Note that $\kappa$ is the conserved spin-orbit coupling quantum number.\cite{sakurai1967aw}} and $V(r)$ is the central potential. For notational simplicity, the subscripts $n\kappa$ on $P$ and $Q$ are omitted throughout this appendix.

The wave functions are normalized according to
\begin{equation}
    \int_0^\infty \big[P^2(r) + Q^2(r)\big] \, dr = 1. \label{eq:normalization}
\end{equation}

Equations (\ref{eq:dpdr}) and (\ref{eq:dqdr}) can be expressed in matrix form as
\begin{equation}
    \frac{d}{dr} 
    \begin{pmatrix}
        P(r) \\ Q(r)
    \end{pmatrix}
    = \mathbf{C}(r)
    \begin{pmatrix}
        P(r) \\ Q(r)
    \end{pmatrix}, \label{eq:dirac_matrix}
\end{equation}
where the coefficient matrix $\mathbf{C}(r)$ is
\begin{equation}
    \mathbf{C}(r) = 
    \begin{pmatrix}
        C_{11}(r) & C_{12}(r) \\
        C_{21}(r) & C_{22}(r)
    \end{pmatrix}, \label{eq:C_matrix}
\end{equation}
with elements (in natural units, $\hbar = c = 1$)
\begin{align}
    C_{11}(r) &= -\frac{\kappa}{r}, \\
    C_{12}(r) &= E - V(r) + m_d, \\
    C_{21}(r) &= -\big[E - V(r) - m_d\big], \\
    C_{22}(r) &= \frac{\kappa}{r}.
\end{align}

For the quark--hadron system, we consider a potential of the form
\begin{equation}
    V_{qH}(r) = \frac{Q_0 \, r^2}{\exp(2r/f_c) - 1} + V_0, \label{eq:potential}
\end{equation}
where $Q_0 \approx 2.4 \times 10^7 \; \text{MeV}^3$
and $V_0 = 4b\sqrt{Q_0}$.
The parameters $V_o$, and $f_c$ are adjusted to reproduce experimental data.

For $r \to 0$, the asymptotic forms of the radial functions depend on the sign of $\kappa$.~\cite{Grant2008} For $\kappa < 0$,
\begin{align}
    P(r) &\sim r^{l+1}, \\
    Q(r) &\sim r^{l+2} \left(\frac{E + Z_1}{2l+3}\right),
\end{align}
and for $\kappa > 0$,
\begin{align}
    P(r) &\sim -r^{l+2} \left(\frac{E + Z_1}{2l+1}\right), \\
    Q(r) &\sim r^{l+1},
\end{align}
where $l$ is the orbital angular momentum quantum number. The constant $Z_1$ represents the $r$-independent part of the potential $V(r)$ evaluated at $r = 0$.

The system of equations (\ref{eq:dirac_matrix}) is solved numerically using a fourth-order Runge--Kutta method for the initial integration steps, followed by an implicit Adams method for the remaining integration. The initial conditions are determined from the asymptotic forms given above, ensuring both regularity at the origin and the number of nodes equal to the principal quantum number $n$. The source code is available upon reasonable request.

 \bibliographystyle{unsrt}

\begin{thebibliography}{99}

 \bibitem{Hsu2025ijmpa} J. P. Hsu and L. Hsu, Int. J. Mod. Phys. A, 2550177 (2025).  
 %(1)

\bibitem{GYM2024}J. P. Hsu and L. Hsu, {\em   General Yang-Mills Symmetry}. World Scientific, 2024. pp. 128-134.
%%%%2  
 
\bibitem{Sonine1880}N. Y. Sonine, Math. Ann.  {\bf 16}, 1-80  (1880).%%%%new7%%%%%
%4%%%3

\bibitem{Hassani1879}M. E. Hassani,  A note on Laguerre original ODEs and Polynomials (1879). Online, Google Scholar.
%%%5%%%4

\bibitem{Hsu2014mpla}J. P. Hsu, Mod. Phys. Lett. A, 2014 {\bf 29}, 1450120, and ref. 2, pp. 132-134.
%(2)%%%6%%%5

\bibitem{Hsu6} J. P. Hsu and L. Hsu, {\em   General Yang-Mills Symmetry}. World Scientific, 2024. pp. 108-113.
%%%6?? ?6
 
\bibitem{Wigner1939}E. Wigner, Ann. Math {\bf 40} 149 (1939).  His classification of particles is based on an irreducible unitary representation of the Poincar\'e group.
 %%%6%%%7
 
%\bibitem{Huang1981ws}K. Huang, {\em Quarks, Leptons \& Gauge Fields}, (World Scientific, 1982)  
%(3)%%%%8
\bibitem{sakurai1967aw}J. J. Sakurai, {\em Advanced Quantum Mechanics}. (Addison-Wesley, 1967), 122-125. %6%%%
%11>>>>>>??????13xxxxx12xxxx8
 
%\bibitem{JJJ88}J. Ahman, et al, Phys. Letts. 3, 206, 364 (1988).
%4%%%9  

%\bibitem{Jaff,95}R. L. Jaffe, Phys. Today, {\bf 48}, 24 (1995)
%%5%%%%10

%xxxxxxxxxx\bibitem{Hsu2024ws}J. P. Hsu and L. Hsu, {\em General Yang-Mills Symmetry, From Quark Confinement to an Antimatter Half-Universe}. World Scientific, 2024. xxii-xxiv, pp. 110-113 for zero energy-momentum of the massless quantum of the `$b_\mu$ fields,' which satisfies the fourth order field equation.  For the derivation of a simple case, i.e., deriving the Hooke potential based on a linear potential, see 132-133. 
%%6%%%%9 %%%11%%%%9
  
 %\bibitem{JPL}preprint?????
%10%%%12

 
\bibitem{Zyla2020ptep}P. A. Zyla {\em et al}, Particle Data Group, Prog. Theor. Exp. Phys., 92-102, 312-324,  (2020).  See 312 for parity of K mesons.%%% 75%
%%12%%%14xxxx9

\bibitem{Navas2024prd}S. Navas  {\em et al}, Particle Data Group, 174-208  (2024).%%%%8%%
%%13%%%%15xxxxx10

%\bibitem{Henry2025thes}H. Raposa, Univ. of Massachusetts Dartmouth, Master Thesis, 'Energy Eigenvalues of a 2-Quark Bound-State System Based on the Relativistic Dirac Hamiltonian with quark-Hooke Potential from Numerical Methods' (2025).%%%9%%
%16
 \bibitem{Grant2008}
I. P. Grant, 
\newblock J. Phys. B. At. Mol. Opt. Phys. {\bf 41} (2008) 1.
%%%%%%10%%15%%%17vvvvv11

%\bibitem{Silbar2010}R. R. Silbar and T. Goldman, Euro J. Phys. 2010, 32, 217(2010).
%%%%%%11%%%%16%%%%18
\bibitem{Workman2022}R.L. Workman et al. (Particle Data Group), Prog.Theor.Exp.Phys. 2022, 083C01 (2022) 
%%%%%%12
\end{thebibliography}

%\end{multicols}
     
 %\clearpage
\end{document}